# Fiber Optical Cable and Connector System (FOCCoS) for PFS/Subaru

Antonio Cesar de Oliveira[1], Lígia Souza de Oliveira[2], Márcio V. de Arruda[1], Lucas Souza Marrara[2], Leandro H. dos Santos[1], Décio Ferreira[1], Jesulino B. dos Santos[1], Josimar A. Rosa[1], Orlando V. Junior[1], Jeferson M. Pereira[1], Bruno Castilho[1], Clemens Gneiding[1], Laerte S. Junior[3], Claudia M. de Oliveira[3], James E. Gunn[4], Akitoshi Ueda[5], Naruhisa Takato[5], Atsushi Shimono[6], Hajime Sugai[6], Hiroshi Karoji[6], Masahiko Kimura[6], Naoyuki Tamura[6], Shiang-Yu Wang[7], Graham Murray[8], David Le Mignant[9], Fabrice Madec[9], Marc Jaquet[9], Sebastien Vives[9], Charlie Fisher[10], David Braun[10], Mark Schwochert[10], Daniel J. Reiley[11]

1- MCT/LNA - Laboratório Nacional de Astrofísica, Itajubá - MG – Brazil.
2- OIO - Oliveira Instrumentação Óptica Ltda. – SP - Brazil
3- IAG/USP – Instituto de Astronomia, Geofísica e Ciências Atmosféricas / Universidade de São Paulo - SP – Brazil
4- Department of Astrophysical Sciences - Princeton University - USA
5- Subaru Telescope/ National Astronomical Observatory of Japan
6 - Kavli Institute for the Physics and Mathematics of the Universe (WPI), The University of Tokyo
7- Institute of Astronomy and Astrophysics, Academia Sinica – Taipei - Taiwan
8- Centre for Advanced Instrumentation, Durham University, Physics Dept.-Durham – UK
9- Observatoire Astronomique de Marseille-Provence, Laboratoire d'Astrophysique de Marseille – Marseille – France
10- Jet Propulsion Laboratory – Pasadena – CA – USA
11- Caltech Optical Observatories – Pasadena – CA - USA

**ABSTRACT**

FOCCoS, "Fiber Optical Cable and Connector System" has the main function of capturing the direct light from the focal plane of Subaru Telescope using optical fibers, each one with a microlens in its tip, and conducting this light through a route containing connectors to a set of four spectrographs. The optical fiber cable is divided in 3 different segments called Cable A, Cable B and Cable C. Multi-fibers connectors assure precise connection among all optical fibers of the segments, providing flexibility for instrument changes. To assure strong and accurate connection, these sets are arranged inside two types of assemblies: the Tower Connector, for connection between Cable C and Cable B; and the Gang Connector, for connection between Cable B and Cable A. Throughput tests were made to evaluate the efficiency of the connections. A lifetime test connection is in progress. Cable C is installed inside the PFI, Prime Focus Instrument, where each fiber tip with a microlens is bonded to the end of the shaft of a 2-stage piezo-electric rotatory motor positioner; this assembly allows each fiber to be placed anywhere within its patrol region, which is 9.5mm diameter.. Each positioner uses a fiber arm to support the ferrule, the microlens, and the optical fiber. 2400 of these assemblies are arranged on a motor bench plate in a hexagonal-closed-packed disposition. All optical fibers from Cable C, protected by tubes, pass through the motors' bench plate, three modular plates and a strain relief box, terminating at the Tower Connector. Cable B is permanently installed at Subaru Telescope structure, as a link between Cable C and Cable A. This cable B starts at the Tower Connector device, placed on a lateral structure of the telescope, and terminates at the Gang Connector device. Cable B will be routed to minimize the compression, torsion and bending caused by the cable weight and telescope motion. In the spectrograph room, Cable A starts at the Gang Connector, crosses a distribution box and terminates in a slit device. Each slit device receives approximately 600 optical fibers, linearly arrayed in a curve for better orientation of the light to the spectrograph collimator mirror. Four sets of Gang Connectors, distribution boxes and Slit devices complete one Cable A. This paper will review the general design of the FOCCoS subsystem, methods used to manufacture the involved devices, and the needed tests results to evaluate the total efficiency of the set.

## 1. INTRODUCTION

The FOCCoS concept consists of 3 different segments of optical fiber cables, which are responsible for guiding light from 2394 positioners to 4 spectrographs. These segments are called, cable A, cable B and cable C. The complete circuit of the subsystem, shown in the layout of Fig.1, is composed of the interposition of several devices like microlenses, fiber arms, protective tubes, optical bench connectors and slit devices. Multi fiber connectors will be used to construct a connector's bench. A tower connector is located between Cable B and cable C; and a Gang connector, between cable B and cable A. The tower connector is necessary to enable removal of PFI from the top end of the telescope. On another hand, the Gang connector is required for logistics reasons.

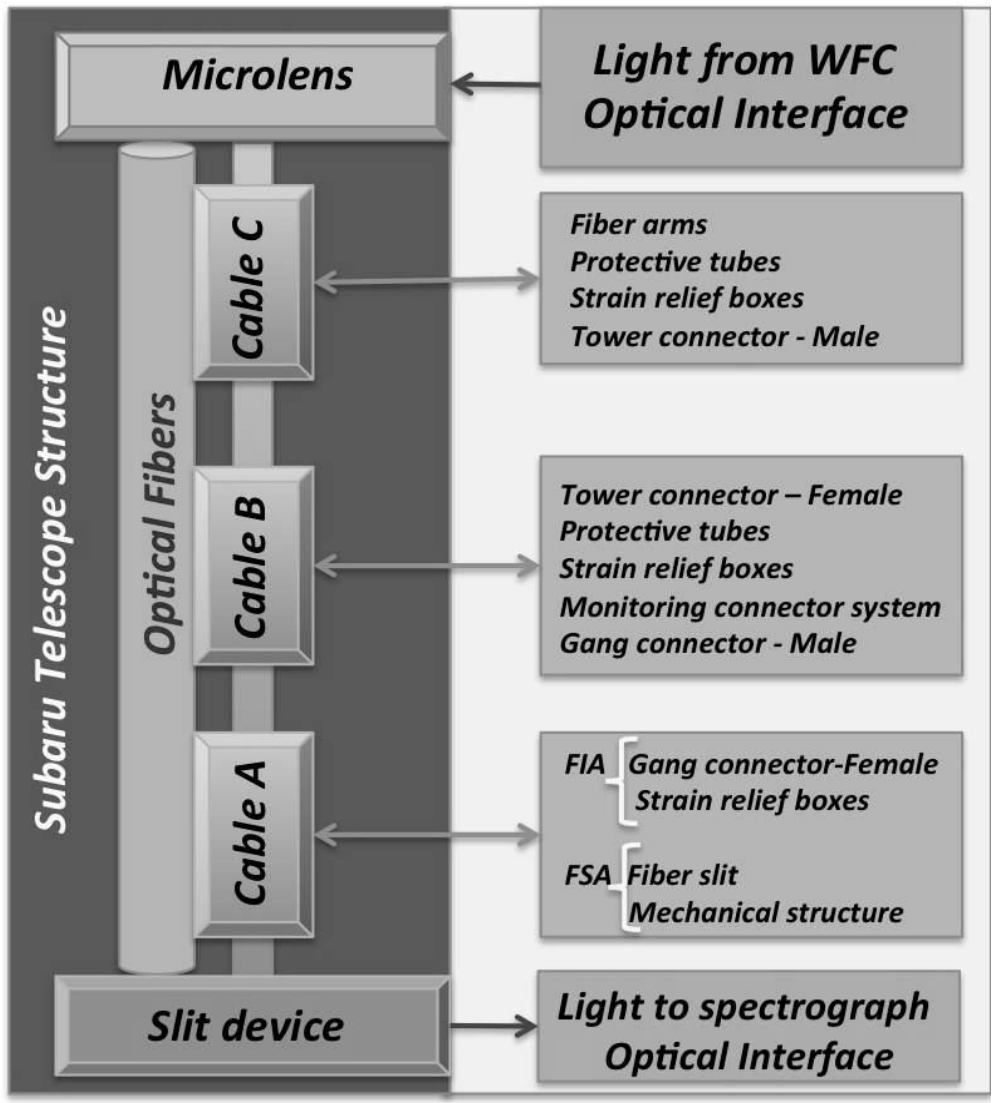

**Figure 1:** FOCCoS Layout subsystem showing the interfaces

FOCCOS subsystem, Figure 1, is defined by 4 interfaces:[01]

1. Telescope (External Physical Interface):
• Cable route & Telescope
The route will be defined taking in account the telescope structure, the available places for the parts of the instrument and the movement of the telescope. (TBD)

2. Spectrograph (External Optical Interface):
• Slit & Spectrograph
For 4 spectrographs we need 4 slits. Each one will be 140mm long and hold 600 fibers, so the center-to-center fiber spacing is 230 microns. The fiber OD is 190 microns.

3. Cobra (External Optical Interface):
• Fiber arms & HSC
Each COBRA positioner unit will contain a fiber arm that holds one fiber to be positioned over a small region of sky. This fiber needs to be polished and secured by COBRA unit in a robust, but stress-free manner, so that no FRD is introduced. The fiber extremity requires a microlens glued to accept the fast (~f/2.3) beam input from the PF corrector with minimal throughput loss.

4. Connectors (Internal Optical Interface):
• Cable A & Cable B & Cable C

Connection points are required for performance of the system with optimal efficiency through a large number of connect/disconnect cycles and the components are designed to avoid deterioration for quality surface of the fibers themselves. Strong connections among all cables; cable A with cable B and cable B with cable C are needed. The connector we considered, which assures this strong connection, is from US Conec, which was already successfully used in the Apogee Spectrograph for the SDSS. All optical interfaces (OI) include a bundle of optical fibers divided in manageable groups. Physical interfaces (PI) refer to all other non-optical interfaces. Each group of optical fibers is protected by plastic protective tubes and clamped accordingly along the route. The route starts at the COBRA positioners, passes through the Tower connectors (84 small units of US Conec connectors), crosses the platform vanes and down, by the telescope tube, passes through the Gang Connectors and finishes at the Spectrograph.

## 2. OPTICAL FIBERS AND PROTECTION TUBES

The FOCCoS system is sequentially divided in Cable A, Cable B and Cable C, which are connected together by multi-fibers connectors. To protect the optical fibers from excessive strain or damage during assembly, tests, integration, installation and operation, several types of protective tubes are specified for use through all the cable routing, as follows:

**2.1- Polyimide tube**

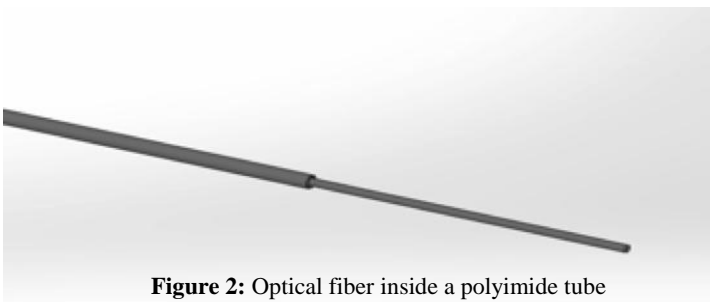

**Figure 2:** Optical fiber inside a polyimide tube

Polyimide tubes are used as strain relief tubes to prevent mechanical stress, and hence FRD, occurring at the point where the fiber enters the ferrule. This tube is assembled on the optical fiber for Cable C immediately after the Zirconia Ferrule ending, as shown in Fig. 2. The nominal specification for the outside diameter is 0.34 mm and for the inside diameter is 0.24 mm.

## 2.2- Segmented tube

Segmented tubes are flexible and made of PTFE plastic. However, the internal surface is extremely smooth, specially designed to accommodate optical fibers. The planned tubes have nominal diameters of 4 mm and 5 mm (outside diameter) and 2,6 mm and 3,1 mm (inside diameter), respectively, as bundles of 33 units of optical fibers can easily be grouped. The external surface is grooved as shown in Figure 3. This kind of tube is good for allowing fibers to spin freely within the tube at the termination points, where longitudinal movement is locked but rotation is allowed. Despite of rotation and elevation movements of telescope structure, torsion on segmented tubes is minimized at grooved area, not stressing the optical fibers. This kind of protection is used for protecting optical fibers in different FOCCoS cables segments, as follows:

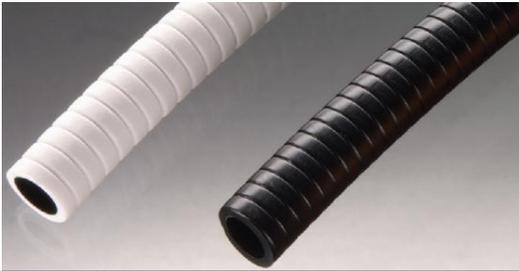

**Figure 3:** Segmented tube with grooving

- Cable C:
  Between Modular Plate # 1 and SRB – Strain Relief Box (IN)
  Between SRB – Strain Relief Box and Tower Connector

- Cable B:
  Between Tower Connector and SRB-I
  Between SRB-I and SRB-II
  Between SRB-II and Gang Connector

- Cable A:
  Between Gang Connector FIA (Fiber Input Assembly) and SRB (Tiny)

## 2.3- Expandable tube

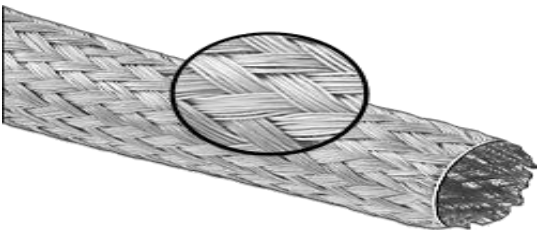

**Figure 4:** Expandable tube

This has the main characteristics of expanding to facilitate installation over a set of segmented tubes and of adjusting itself around its content. This tube is double braided to provide flexibility and provides good resistance to abrasion. Figure 4 shows the expandable tube geometry. The segmented tubes between SRB and Tower connector, in Cable C, will be bundled in groups of 21 units and will be protected by one unit of expandable tube.

## 2.4-Supporting Tube

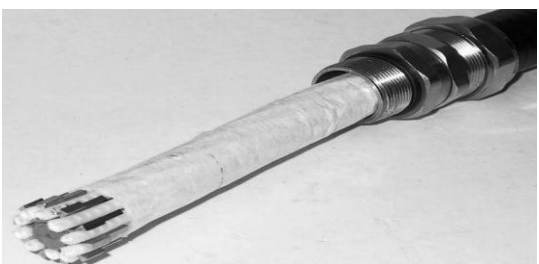

**Figure 5:** Supporting tube

Supporting tubes have the main function of holding Cable B, mainly at its longest portion between SRB-I and SRB-II, where Cable B is under tension due to gravity and its vertical positioning. To support this force, this tube is manufactured with an external tube of plastic, an internal tube of PTFE and, between them, a reinforcement made of Kevlar fibers, as shown in Figure 5. A set of 21-segmented tubes is bundled and fixed by an adhesive tape around one Supporting tube. The Kevlar reinforcement tip is prepared as it can be fastened, independently and tighter than segmented tubes, as to receive and to resist the tension forces.

# 3 – CABLE SYSTEM AND FOCCoS DESIGN

The basic design of the FOCCoS sub-system is premised on the use of 3 fiber optic cables connected by multi-fiber connectors. The cables that make up the entrance and exit the system are respectively optimized to capture light from the telescope and inject light into the spectrographs. The central cable is the light conductor between the extremities.

## 3.1 - Cable A

Cable A is the cable installed at the Spectrograph side and consists of the FSA (Fiber Slit Assembly), the routing and its support, and the FIA (Fiber Input Assembly), as shown in Fig. 6. Cable A is composed of a set of optical fibers arranged linearly on the slit component and supported by the Frame, protected by segmented tubes and routed between strain relief boxes, and the connection interface at the Support Connector (*Gang*) to allow connection with Cable B, at the Subaru Telescope interface. As four Spectrographs are considered for PFS/Subaru, four units of Cable A are necessary.

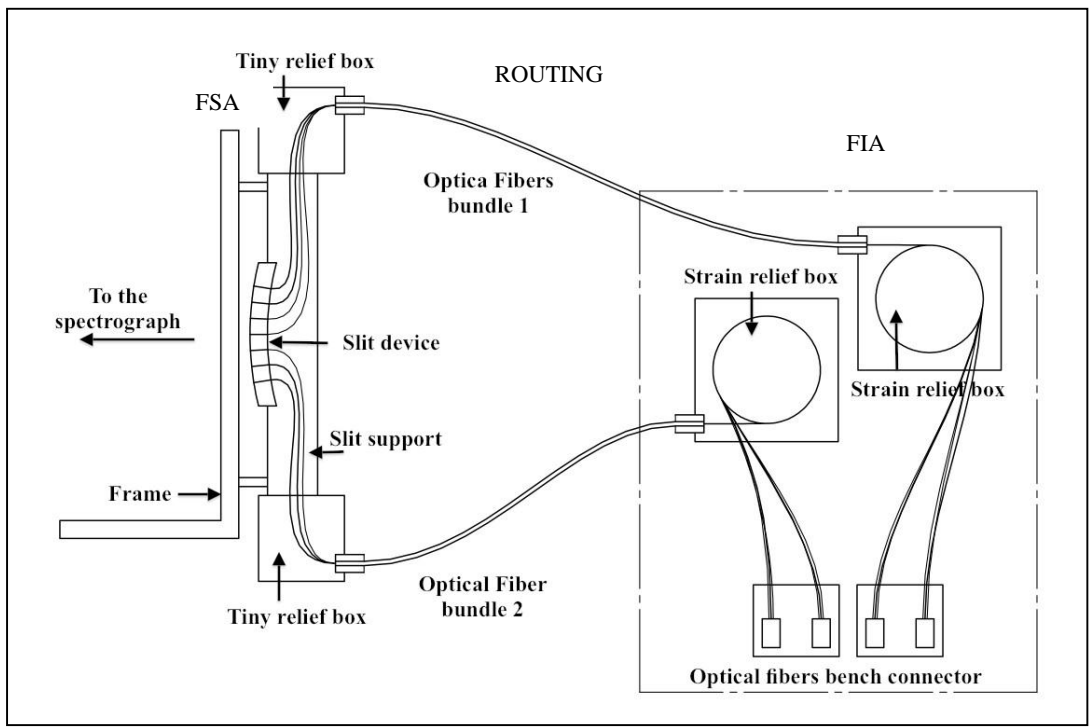

**Figure 6:** Schematic view of Cable A

The Fiber Slit Assembly, FIA, ensures proper positioning of the fibers by using masks of micro-holes. This kind of mask is made by a technique called electro-formation, which is able to produce a nickel plate with holes in a precise linear array. The precision error is around 1 μm for diameter and 1 μm for pitch position of the holes. This nickel plate may be produced with a thickness between 50 μm and 200 μm, so it can be very flexible. This flexibility allows inducing the curvature necessary to match the curved surface found on the slit. This complex geometry is implemented using two masks, called Front Mask and Rear Mask, which are separated by a gap that defines the thickness of the slit. In this context, the pitch and the diameter of the holes define the linear geometry of the slit. On the other hand, the curvature radius of each mask defines the angular geometry of the slit. Slit devices, as a part of terminations of optical fibers suitable to be used inside a Schmidt camera, are subject to dimensional limitations due to the shadow generation inside the optical beam propagated along the camera. The size limitation on the slit, if not considered during the project, may cause structural and spatial problems whose effects are felt in the efficiency of the system. This concept presents several advantages if compared with the classical options. It is much easier to assemble in a short period of time, much cheaper, accurate, easy to adjust; it also offers the possibility of making a device, which is much stronger, robust (and completely miniaturized) than the conventional option. Obviously to be supported inside the spectrograph, this set needs to be assembled inside some structure, rigid and strong enough. Furthermore, all structure needs to have a structural CTE optimized to avoid problems of displacement of the fibers or increase of FRD in the fibers when the device is submitted to the low temperature of operation for the spectrograph (3 °C). A possible solution is presented in Fig. 7, which shows a frame with the slit support and the slit device. The zoomed figure on the right shows a part of the Schmidt camera inside the spectrograph. The material used is Titanium and INVAR to reduce weight and provide a good CTE combination.

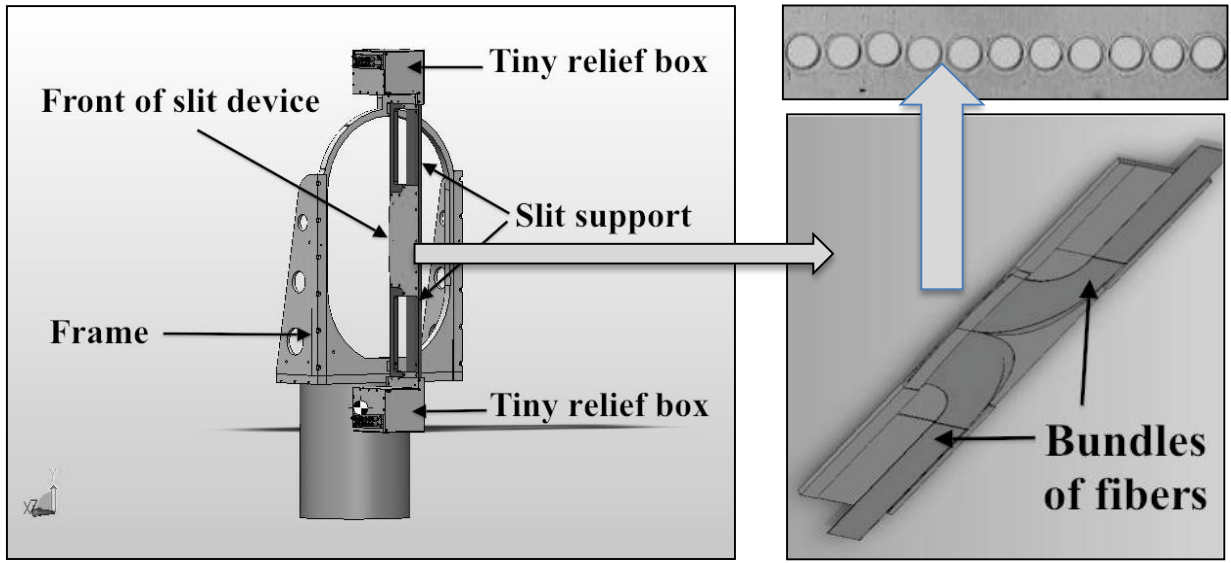

**Figure 7:** Fiber Slit Assembly, FSA, with two small strain relief boxes necessary to accommodate both combs of optical fibers from the slit. The slit support thickness is between 7 and 8 mm to avoid light loss, due to the shadow, naturally created inside the Smith camera.

The optical fiber Support Connectors *(Gang connectors)* are the other extremity of Cable A and are responsible for optical connection with Cable B. This connector's support is called Gang Connector and is shown in Fig.8. Each Connector Support has 12 units of multi-fibers Usconec Ferrule 32F. [02]

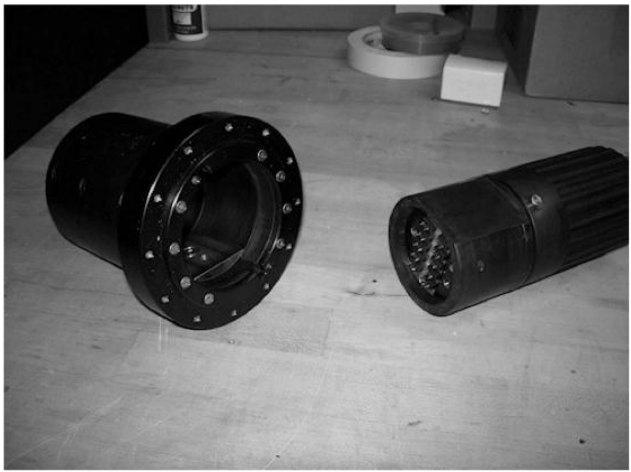

**Figure 8:** Gang Connector device

Each Cable A starts at the Support Connector, and is then divided into two units which form two bundles of optical fibers, until they reach the Strain Relief Boxes where other bundles of optical fibers are formed, now protected by segmented tubes. These bundles are disassembled inside Tiny SRB and the optical fibers are linearly arranged on Slit curved surface, finishing Cable A. The Gang Connector and SRBs are installed in an especially designed support bench, called Fiber Input Assembly (FIA), placed in the IR fourth floor in the Subaru Telescope building. The length of cable A is mainly defined by the available space for the four FIA's units and by the distribution of the Spectrographs' benches in the room. Inside the Spectrographs, the slit devices are mounted on the so-called Fiber Slit Assembly (FSA).   The FSA is a mechanical support, designed to be over a dithering mechanism that will help with the calibration of the instrument. The FSA includes indexed mounting features to facilitate removal and reinstallation from/to the instrument without requiring re-alignment. A schematic view of one unit of Cable A is shown in Fig. 9.

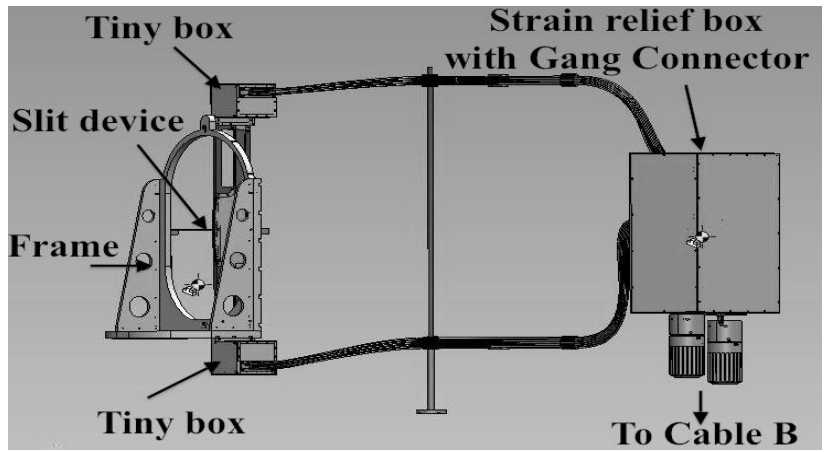

**Figure 9:** Schematic of one Cable A with FIA, Fiber Slit Assembly, and the Gang Connector bench between the extremities

### 3.2 Cable B

Cable B is the cable installed at the Subaru Telescope structure and consists of specific supports for connectors at its sides, where there are the interfaces with Cable A and Cable B, as shown in Fig. 10.

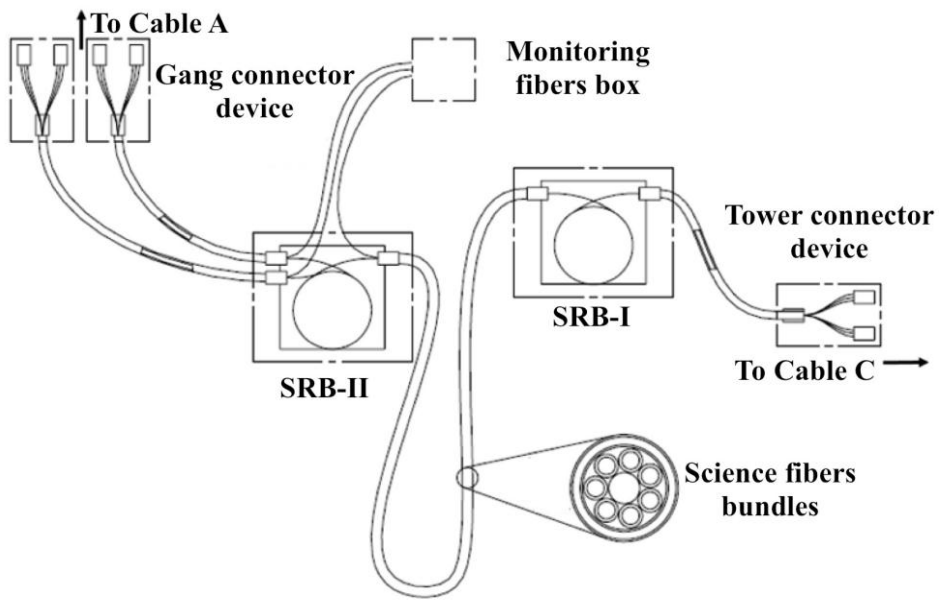

**Figure 10:** Schematic view of Cable B.

Cable B, like in FMOS instrument[02], is the cable permanently installed at the telescope structure, composing the longest cable of the system, around 57 meters. This cable is, basically, a conduit tube containing several plastic tubes inside (segmented tubes) through which the groups of optical fibers are protected. The routing of the cable, Fig.11, is either looped across to the "Great Wall", within which the spectrographs are housed, or through the cable wrap of the elevation axis. The cable interfaces directly with the telescope and dome structure from the TES vanes to the Great Wall. It must minimize stress on the fibers during telescope pointing and field rotation of the TES.

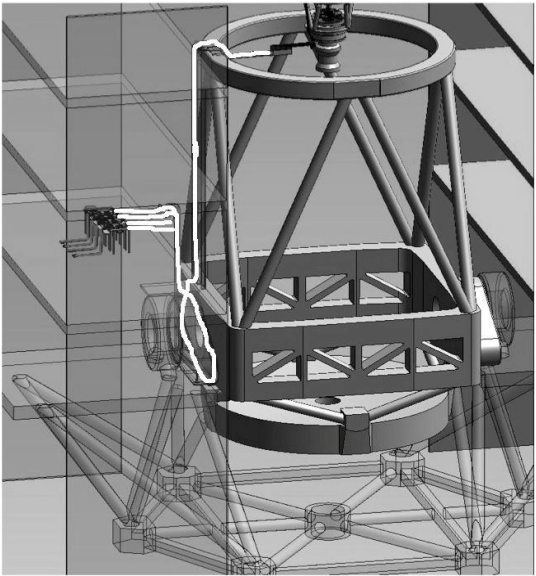

**Figure 11:** Possible route for Cable B, shown by the white line

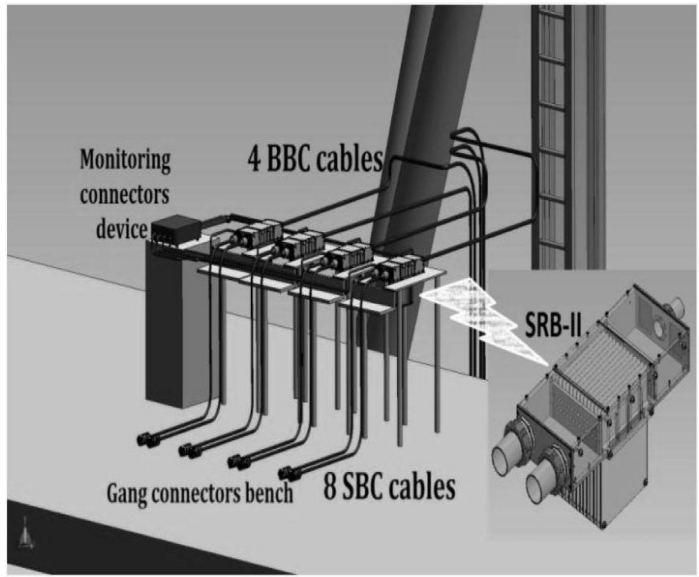

**Figure 12:** Cable B termination in the Strain Relief Box, Gang connector's male and the Monitoring Connection System.

Changes in the length between local fiber securing points will be accommodated by constant force springs. Strain relief boxes, Fig.12, are located at both the extremities of cable B: close to the spectrograph system, and close to the top end system. This separation defines enough free length of fibers out of the protection tubes to facilitate the manipulation of the fibers during the construction and the polishing procedures of the cable. Cable B, at the top end side, is separated in eight cables within segmented tubes to conduct the optical fibers. Figs. 12 and 13 show details of the Strain relief Boxes and a possible position to be installed at the spectrograph room. The Cables for this purpose will be made following helicoid format structure. The cables are divided into two parts; Big Black Cable (more or less 90% the route) and Small black Cables 10% route. The SRB will have 2 loops of optical fiber.

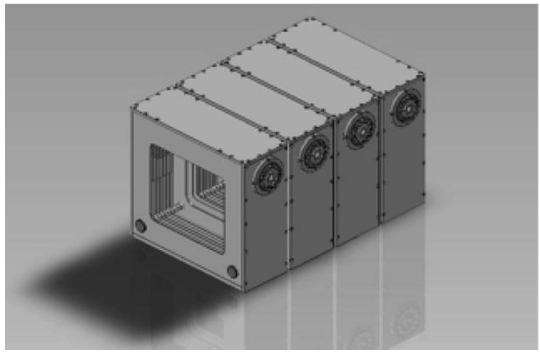

**Figure 13:** Strain Relief Box, SRB I, that will be fixed at the top end of the telescope.

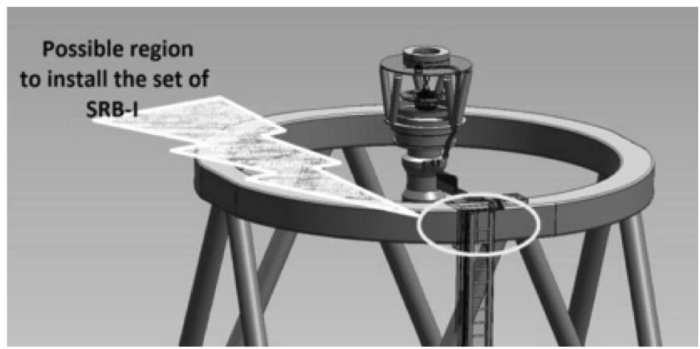

**Figure 14:** The Cable B, starts at the top end. This cable is a set of fours sub-cables passing through four sets of SRBs.

## 3.3 Cable C

Cable C is permanently installed inside the Prime Focus Instrument (PFI), at the top of Subaru Telescope. Its main function is to capture the skylight and to couple this light into Cable B. A schematic view of Cable C is showed in Fig. 15.

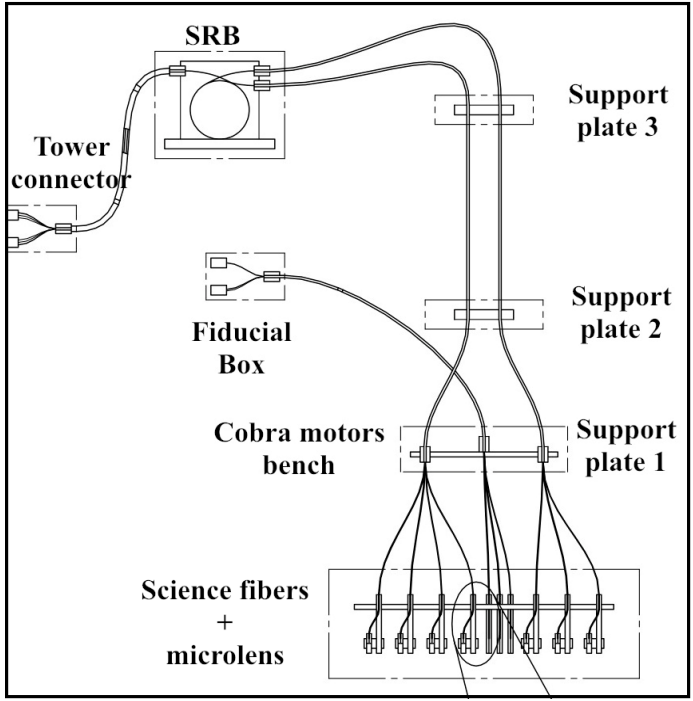

**Figure 15:** Schematic of Cable C

The light, reflected by the telescope mirror, reaches each one of the 2400 optical fibers by individual microlenses placed on its extremity. Each fiber is assembled at the shaft, which is the extremity of a 2-stage piezo-electric rotatory motor positioner that is able to point each optical fiber within its patrol region of diameter 9.5 mm, as a small region of sky can be scanned. Each fiber extremity, inserted inside a ferrule, is supported in each positioner by a fiber arm device, Fig. 16, coupled with a microlens to collect light from WFC. The geometry of rays coming from the WFC defines a cone of light with focal ratio F/2.19 at the center of the field of view and around F/2.01 at its edge, because of the non-telecentricity of the system. Clearly, this is a problem for the fibers that has NA=0.22, which implies in an equivalent limit of focal ratio F/2.27. To increase the F-ratio after the wide-field corrector by a factor of 1.28, a microlens will be cemented at the extremity of each fiber in the fiber arm. The current design of the microlens and it specifications are shown in Fig. 17. Accordingly, the F-ratio changes as follows: Field center: F/2.19 ↳ F/2.8; Field edge: F/2.01 ↳ F/2.57.

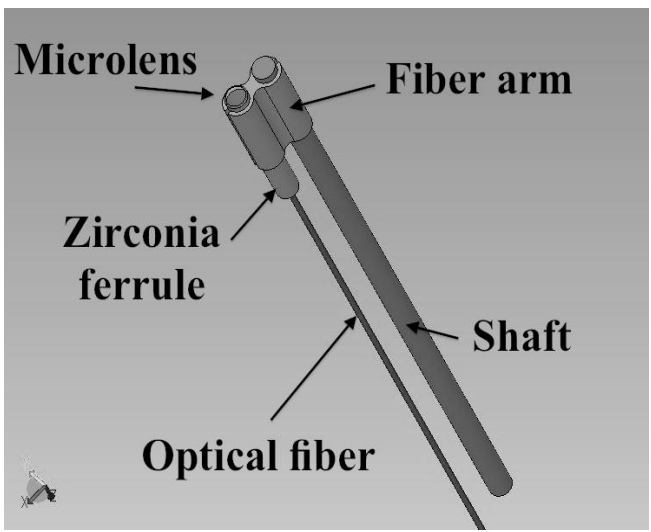

**Figure 16:** View of the fiber arm set, showing in details: microlens, fiber arm, zirconia ferrule, and the steel shaft.

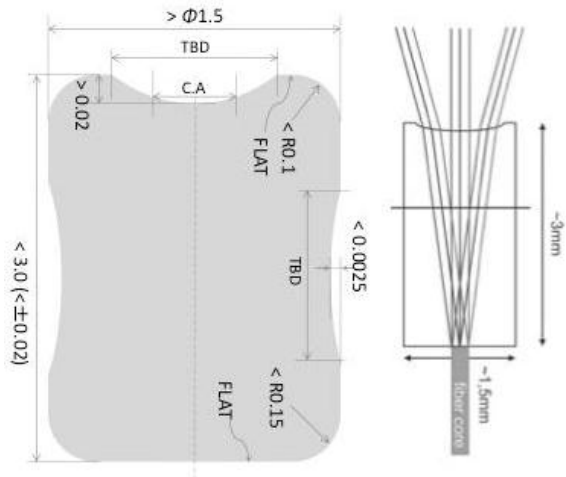

**Figure 17:** Lens magnification of x 1.28 is at λ = 0.76 μm.

Special care is necessary during the polishing process of the ferrule and during the assembly onto the optical fiber. The flatness and roughness are must be controlled to meet system requirements for low FRD and high throughput. The 2400 optical fibers with their motor positioners are assembled on a bench plate, arranged in a hexagonal-closed-packed pattern, as shown in Fig. 18. The other ends of these fibers will be terminated in the Tower Connector. All optical fibers from Cable C, protected by segmented tubes, are routed through the motors' bench plate and three sequential modular plates before reaching a strain relief box, as shown in Fig. 19. The individual position for each optical fiber is addressed in accordance with a fiber's mapping to the spectrograph; its route can be found and identified at the connector's interface as well as at the slit device.

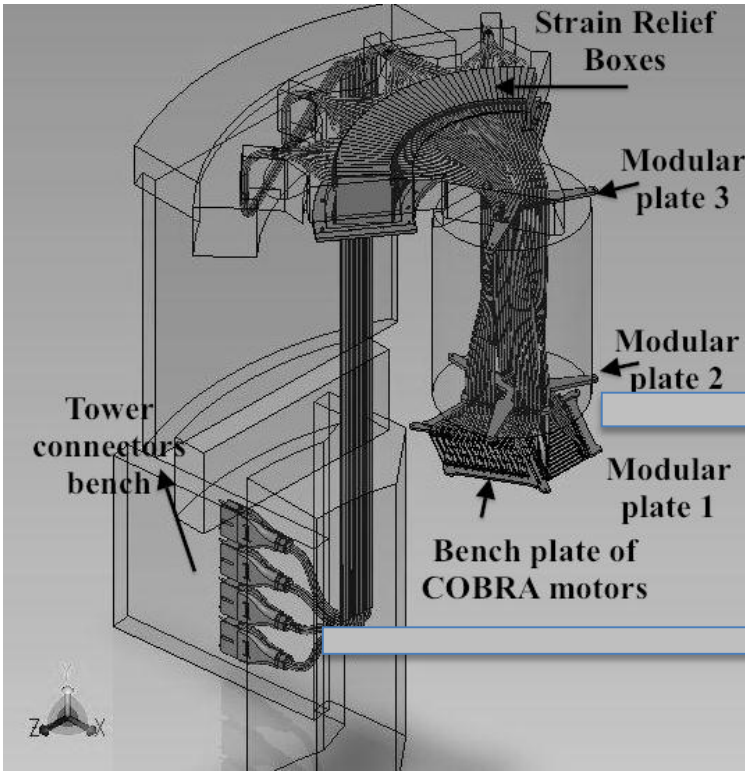
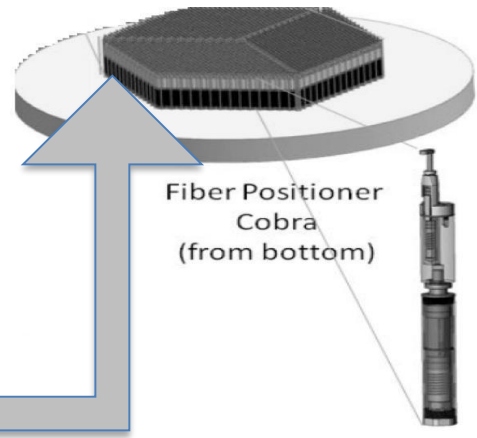

**Figure 18:** Bench plate with 2400 optical fiber positioners

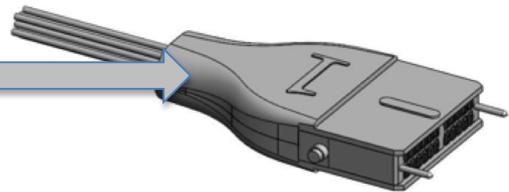

**Figure 19:** Routing for Cable C from Modular Plate 1 to SRB

**Figure 20:** Terminal part of the Cable C

After the SRB, the optical fibers are arranged in bundles of 11- segmented tubes and are routed to the Tower Connector, which is the bench to support the US Conec multi-fibers connectors. A strong connection force with Cable B is needed at the Tower Connector. When the PFI instrument is removed, the Tower Connector is disassembled and parked in a specific position on the Cable wrapper area. Fig. 20 shows the schematic view of the final portion of Cable C (tower connector).

**3.4 Optical fibers**

The FOCCoS sub-system uses optical fibers from two different manufacturers, Fujikura and Polymicro; it is comprised of three fiber optic cables coupled together by multi fibers connectors. The fiber geometry is about 128-μm diameter core, 170-μm diameter cladding, and 190 μm diameter buffer. The optical fibers of each cable have slightly different core diameters such that the light is conducted from smaller cores to larger cores, thereby improving the performance of connections. Both types of fibers require a robust structure with large diameter of cladding and buffer to support the effects of torsion, which is caused the fiber positioners. The optical fibers tested in this work were the FBP Polymicro and BPI Fujikura. Table I, shows the established combinations for core diameter and length of the cable.

Table I: Samples in test for this work

| Company | Cable A | Cable B | Cable C |
|---|---|---|---|
| Fujikura | | 55 meters: S.128/170 BPI | |
| Polymicro | 2 meters: FBP 129168190 | | 7 meters: FBP 127165190 |

## 4. SUMMARY AND CONCLUSIONS

We described here the design for FOCCoS, (Fiber Optical Cable and Connectors System) to be a subsystem of SuMIRe-PFS, (SUBARU Measurement of Images and Redshifts- Prime Focus Spectrograph). A set of 4 Slits will comprise one of the extremities of the cable system. The other extremity will direct the fibers for fiber arms devices as the part of a patrol system. Each fiber arm will have one end of optical fiber polished and coupled at a microlens. FOCCoS is will work with 2394 optical fibers segmented in 3 cables connected by sets of multi-fibers connectors. The multi-fiber connectors that will be the used is produced by USCONEC. Polymicro and Fujikura produce the optical fibers actually in test. Polymicro fibers will be used in Cable A and Cable C; Fujikura fibers will be used in Cable B.

## 5. ACKNOWLEDGEMENTS

We gratefully acknowledge support from: The Funding Program for World-Leading Innovative R&D on Science and Technology. SUBARU Measurements of Images and Redshifts (SuMIRE), CSTP, Japan. Fundação de Amparo a Pesquisa do Estado de São Paulo (FAPESP), Brasil. Laboratório Nacional de Astrofísica, (LNA) e Ministério da Ciência Tecnologia e Inovação, (MCTI), Brasil. We would like also to gratefully acknowledge Youichi Ohyama and Hung-Hsu Ling, PO members.

## 6. REFERENCES

[01] Sugai, H., et al., "Prime focus spectrograph: Subaru's future" Proc. SPIE 8446, Ground-based and Airborne Instrumentation for Astronomy IV, 84460Y (2012)
[02] Murray, G. J., Dodsworth, G. N., Content, R., Tamura, N., "Design and construction of the fiber system for FMOS" Proc. SPIE 7014, 70145L (2008).
[03] Brunner, S., et al., "APOGEE fiber development and FRD testing" Proc. SPIE 7735, Ground-based and Airborne Instrumentation for Astronomy III, 77356A (2010)